\pdfoutput=1
 \documentclass[aps,prd,reprint,twocolumn,showpacs,superscriptaddress,nofootinbib]{revtex4-1}  

\usepackage{tabularx}
\makeatletter
\def\hlinewd#1{%
\noalign{\ifnum0=`}\fi\hrule \@height #1 %
\futurelet\reserved@a\@xhline}
\makeatother
 \usepackage{auncial}
\usepackage{scrextend}
\usepackage{relsize}
\usepackage{amsmath}
\usepackage{amssymb}
\usepackage{epsfig}
\usepackage{graphicx}
\usepackage{hyperref}
\usepackage{dcolumn}   
\usepackage{tabu}
\usepackage{boldline}
\usepackage{slashed}
\usepackage{multirow}
\usepackage{bbold}
\usepackage{color}
\usepackage[normal]{subfigure}
\usepackage{rotating}
\usepackage[margin=0.9in,a4paper]{geometry}
\usepackage[table]{xcolor}
\usepackage{enumitem}
\usepackage[normalem]{ulem}
\usepackage{colortbl}
\usepackage{array,multirow}
\definecolor{nicered}{rgb}{0.7,0.1,0.1}
\definecolor{nicegreen}{rgb}{0.1,0.5,0.1}
\definecolor{red}{rgb}{1.0, 0, 0}

\allowdisplaybreaks

\definecolor{Grn}{rgb}{0.,0.75,0.}
\definecolor{Blu}{rgb}{0.,0.,1.}

\def\gsim{\raise0.3ex\hbox{$\;>$\kern-0.75em\raise-1.1ex\hbox{$\sim\;$}}}
\def\lsim{\raise0.3ex\hbox{$\;<$\kern-0.75em\raise-1.1ex\hbox{$\sim\;$}}}

\def\mb[#1]{\mathbf{#1}}

\renewcommand{\bar}{\overline}

\definecolor{LightCyan}{rgb}{0.88,1,1}
\definecolor{piggypink}{rgb}{0.99, 0.87, 0.9}
\definecolor{applegreen}{rgb}{0.55, 0.71, 0.0}
\definecolor{darkpastelgreen}{rgb}{0.01, 0.75, 0.24}
\definecolor{green-yellow}{rgb}{0.68, 1.0, 0.18}

\newcommand{\beq}{\begin{equation}}
\newcommand{\eeq}{\end{equation}}
\newcommand{\beqa}{\begin{eqnarray}}
\newcommand{\eeqa}{\end{eqnarray}}

\newcommand{\g}{\gamma}

\newcommand{\bi}{\begin{itemize} }
\newcommand{\ei}{\end{itemize} }

\begin{document}



\title{Shining dark matter in Xenon1T}

\author{Gil Paz}
\affiliation{\normalsize \it 
Department of Physics and Astronomy,
Wayne State University, Detroit, Michigan 48201, USA
}

\author{Alexey A. Petrov}
\affiliation{\normalsize \it 
Department of Physics and Astronomy,
Wayne State University, Detroit, Michigan 48201, USA
}

\author{Michele Tammaro}
\affiliation{\normalsize \it 
Department of Physics, University of Cincinnati, Cincinnati, Ohio 45221,USA
}

\author{Jure Zupan}
\affiliation{\normalsize \it 
Department of Physics, University of Cincinnati, Cincinnati, Ohio 45221,USA
}


\begin{abstract}
\noindent We point out that a non-relativistic $\sim 2 $ GeV dark matter (DM) which interacts with visible matter through higher dimensional Rayleigh operators could explain the excess of  ``electron recoil'' events recently observed by the Xenon1T collaboration. A DM scattering event results in a few keV photon that on average carries most of the deposited energy, while the nuclear recoil energy is only a subleading correction. Since the Xenon1T detector does not discriminate between electrons and photons, such events would be interpreted as excess of the keV electrons. Indirect constraints from dark matter annihilation are avoided for light mediators of ${\mathcal O}(10~{\rm MeV})$ that have sizable couplings to neutrinos. One loop induced spin-independent scattering in dark matter may soon lead to a confirmation signal or already excludes regions of viable parameter space for the Rayleigh DM model, depending on what the exact values of the unknown nonperturbative nuclear matrix elements are. 
\end{abstract}

\maketitle

\flushbottom

{\bf Introduction.} Xenon1T collaboration recently announced the results of a search for Dark Matter (DM) using electronic recoils after 0.65 tonne-years of exposure \cite{Aprile:2020tmw}. An anomalously large number of events were observed as a peak over the nominal background at the threshold of the experimental sensitivity. Two different explanations were proposed to date. Firstly, the tritium induced background may be significantly enhanced well above the Xenon1T estimate. Secondly, this could be the first signal of new physics that interacts predominantly with electrons.

For instance, when interpreted as an absorption of a solar axion, the excess events correspond to a $3.5\sigma$ deviation over the background only hypothesis. The significance is somewhat reduced if it is interpreted as  solar neutrinos scattering on electrons via nonzero neutrino magnetic moment \cite{Aprile:2020tmw}, or as the absorption of bosonic dark matter/ALP on electrons \cite{Aprile:2020tmw,Takahashi:2020bpq}. Stellar cooling bounds are in tension with the solar axion \cite{Giannotti:2017hny,Diaz:2019kim} and neutrino magnetic moment intepretations~\cite{Corsico:2014mpa,Diaz:2019kim,Bell:2006wi,Bell:2005kz}, but not with bosonic dark matter \cite{Calibbi:2020jvd}.   A number of alternative explanations were also proposed: fast moving DM particles scattering on electrons \cite{Fornal:2020npv,Kannike:2020agf}, nonstandard neutrino interactions mediated by light new particles \cite{Boehm:2020ltd}, hidden photon dark matter \cite{Alonso-Alvarez:2020cdv}, etc.

In this Letter we propose a third possibility, 
that the anomalous events are due to electromagnetic interactions of non-relativistic DM with xenon {\it nuclei}. That this is a realistic possibility is somewhat surprising.
Firstly, electromagnetic interactions lead to scatterings on both nuclei and electrons. Secondly, for non-relavistic DM only the scatterings on nuclei result in  large enough energy transfers, of a few keV, so that these can be observed in the Xenon1T detector. 
In contrast, the observed excess events are unmistakably of the ``electron recoil'' type (energy deposited in photons and/or electrons), which seems to rule out the possibility of elastic scatterings of non-relativistic DM. 

\begin{figure}[t]
 \includegraphics[width=3.8cm]{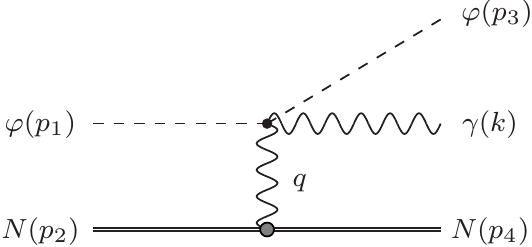}
 ~ 
 \includegraphics[width=3.8cm]{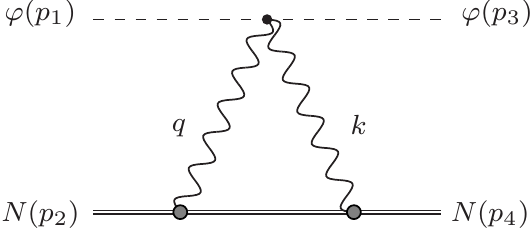} 
\caption{Left: The tree level diagram for $\varphi N\to \varphi N \gamma$ scattering for Rayleigh DM. Right: The one-loop diagram contributing to $\varphi N\to \varphi N$ scattering.
\label{fig:diagram}}
\end{figure}

\begin{figure}[t]
 \includegraphics[width=6.cm]{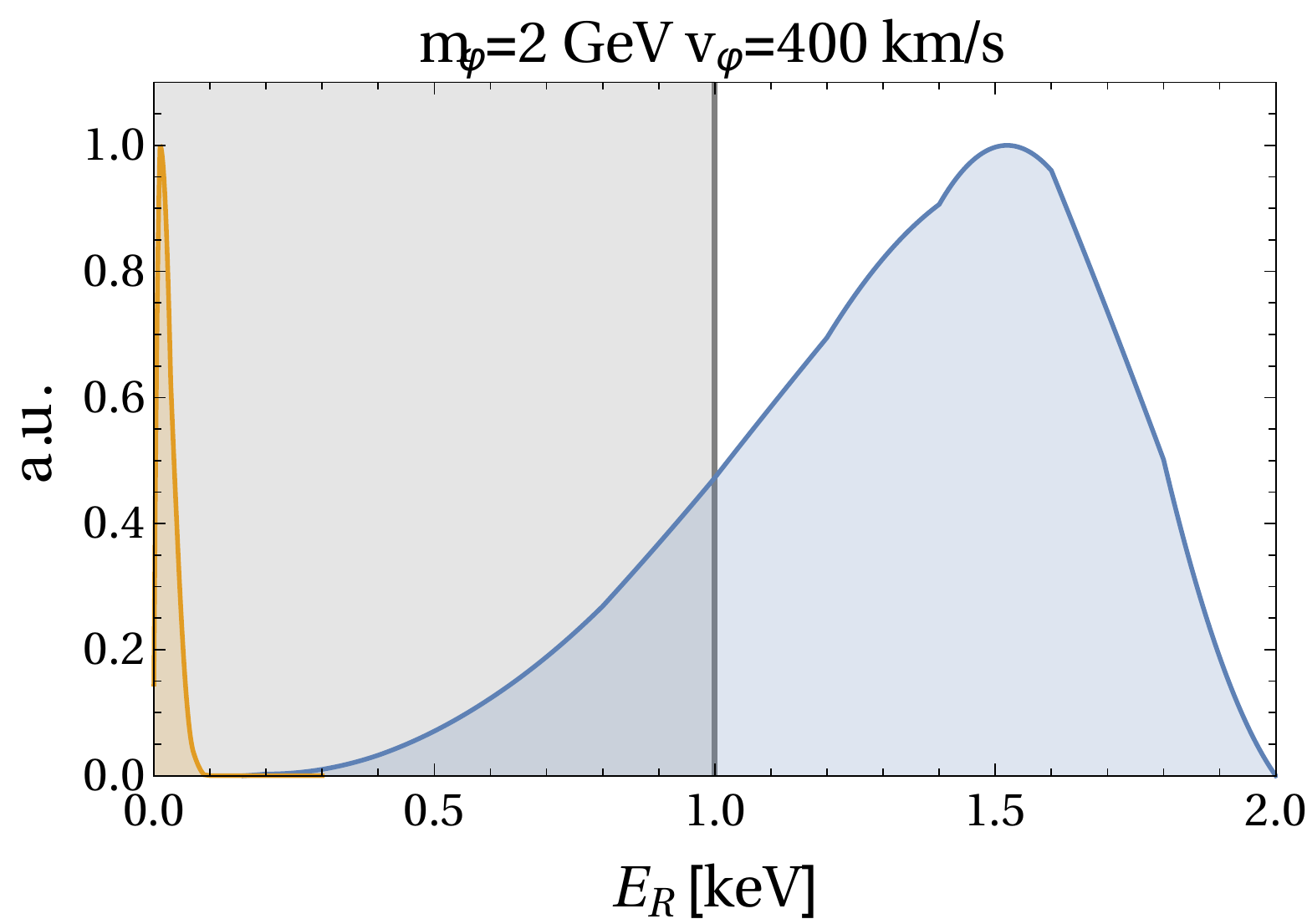} 
\caption{The simulated recoil energy spectra, $E_R=\{E_\gamma, E_{\rm NR}\}$, in the $\varphi A\to \varphi A \gamma$ scattering through the CP conserving Rayleigh operator, Eq. \eqref{eq:Rayleigh:scalar}, for a monochromatic DM of fixed incoming velocity $v = 400$ km/s and mass $m_\varphi=2$ GeV. The energy spectrum of the emitted photon (blue shaded region) is significantly harder than the energy due to the nuclear recoil (orange shaded region). The gray shading indicates the 1 keV detector threshold for electron recoils. The distributions were obtained using {\tt Feynrules} \cite{Christensen:2008py} and {\tt Madgraph5} \cite{Alwall:2011uj}. }
	\label{fig:recoil400}
\end{figure}

The exception to this naive conclusion is DM that couples to the visible sector through the Rayleigh operator, $\varphi\varphi F_{\mu\nu}F^{\mu\nu}$. In this case the  $2\to3$ scattering on a xenon nucleus $N$, $\varphi N\to \varphi N\gamma$, is possible, as shown in Fig. \ref{fig:diagram} (left). The majority of the deposited energy is carried away by the  photon, while the nuclear recoil energy is much smaller, cf. Fig. \ref{fig:recoil400}. To a very good approximation these events are indistinguishable from the pure electron recoil events.  

At one-loop the Rayleigh operator also induces spin-independent (SI) $\varphi N\to \varphi N$ scattering \cite{Weiner:2012cb,Frandsen:2012db}, see Fig. \ref{fig:diagram} (right). As we explain below,
this gives important constraints on the model. 

{\bf Rayleigh dark matter.} For concreteness we assume that DM is a real scalar, $\varphi$, which couples to the 
visible sector through dimension-6 Rayleigh operators,
\beq
\begin{split}\label{eq:Rayleigh:scalar}
{\cal L}_{\rm int}=\frac{\alpha}{12\pi}  \frac{1}{\Lambda^2} \varphi^2 \Big(& C_{\gamma} F_{\mu\nu} F^{\mu\nu}
 + \widetilde C_{\gamma} F_{\mu\nu} \widetilde F^{\mu\nu}\Big),
\end{split}
\eeq
where $F_{\mu\nu}$ is the electromagnetic field strength.  DM is assumed to be $Z_2$-odd and thus stable, while the SM fields are $Z_2$-even. The first (second) operator in 
Eq.~\eqref{eq:Rayleigh:scalar} is CP conserving (violating).

The Rayleigh operators may well be the leading interactions between the SM and the dark sector \cite{Kavanagh:2018xeh,Weiner:2012gm}. For instance, $C_\gamma, \tilde C_\gamma\sim {\mathcal O}(1)$ are generated at one loop, if DM couples to heavier states of mass ${\mathcal O}(\Lambda)$  charged under the SM electroweak group~\cite{Weiner:2012gm,Fichet:2016clq,Weiner:2012gm}.
For Dirac fermion DM, the one-loop radiative corrections generically also induce the DM magnetic moment.
In contrast, for real scalar DM or Majorana fermion DM the operators of lowest  dimension that couple DM to gauge bosons are, in fact, the Rayleigh operators.

{\bf Signature in Xenon1T.} The direct detection signatures of Rayleigh DM are of two types: i) a purely nuclear recoil event $\varphi N\to \varphi N$ induced at one loop through two photon exchange, and ii) the $\varphi N\to \varphi N \gamma $ scattering, in which the energy is distributed between the nuclear recoil and the energy of the photon.
The cross section for the $\varphi N\to \varphi N \gamma $ scattering is given by 
\beq
\frac{d\sigma}{dE_{\rm NR} dE_\gamma}=\frac{1}{16}\frac{1}{(2\pi)^3}\frac{\big|{\cal M}\big|^2}{m_\varphi m_N v},
\eeq
where $E_{\rm NR}$ is the recoil energy of the nucleus, $E_{\gamma}$ the photon energy, $m_\varphi$ and $m_N$ are, respectively, the masses of the DM and of the nucleus, and $v\sim 10^{-3}$ the velocity of the incoming DM. We work in the nonrelativistic limit, assuming $m_\varphi\ll m_N$, so that the lab frame coincides with the center of mass frame for the scattering. The matrix element squared is given by 
\beq
\begin{split}
\big|{\cal M}\big|^2&=   \biggr(\frac{2 \sqrt 2}{3\pi}\frac{\alpha Z e C_\gamma}{\Lambda^2}\biggr)^2\frac{1}{(Q^2)^2}\Big[ Q^2\times
\\
&  \times \big((k\cdot p_2)^2+(k\cdot p_4)^2\big)- 2 m_N^2 \big(k\cdot q)^2\Big],
\end{split}
\eeq
where $q^\mu=p_4^\mu-p_2^\mu$, and $Q^2\equiv-q^2=2m_N E_{\rm NR}$, with the four momenta as defined in Fig. \ref{fig:diagram} (left), and $Z=54$ the atomic number of xenon. 
The differential cross section peaks toward small values of $E_{\rm NR}$ due to the photon pole and the large mass of the nucleus, while the emitted photon tends to have the maximal energy, see Fig. \ref{fig:recoil400}. Here and below we set $\widetilde C_\gamma=0$. However, all our results apply also to the CP violating case, with $C_\gamma\to \widetilde C_\gamma$ replacements.

The signal rate in the Xenon1T detector is given by
$
{dR}/{dE_\gamma}={\rho_0}\int_{v>v_{\rm min}}d^3 v \, v f_\odot (\vec v) ({d\sigma}/{dE_\gamma}) /({m_\varphi}{m_N}),
$
where $v_{\rm min}\simeq \sqrt{2 E_\gamma/m_\varphi}$ up to small corrections of ${\mathcal O}(E_\gamma^2/m_Nm_\varphi)$, and $\rho_0=0.3~{\rm GeV}/{\rm cm}^3$ is the local DM density. For the DM velocity distribution $f_\odot(v)$ we use the standard model halo type distribution
 truncated at the escape velocity $v_{\rm esc}=550$ km/s, and width of $\bar v=220$ km/s, see, e.g., \cite{Fairbairn:2008gz}.

\begin{figure}[t]
 \includegraphics[width=6.5cm]{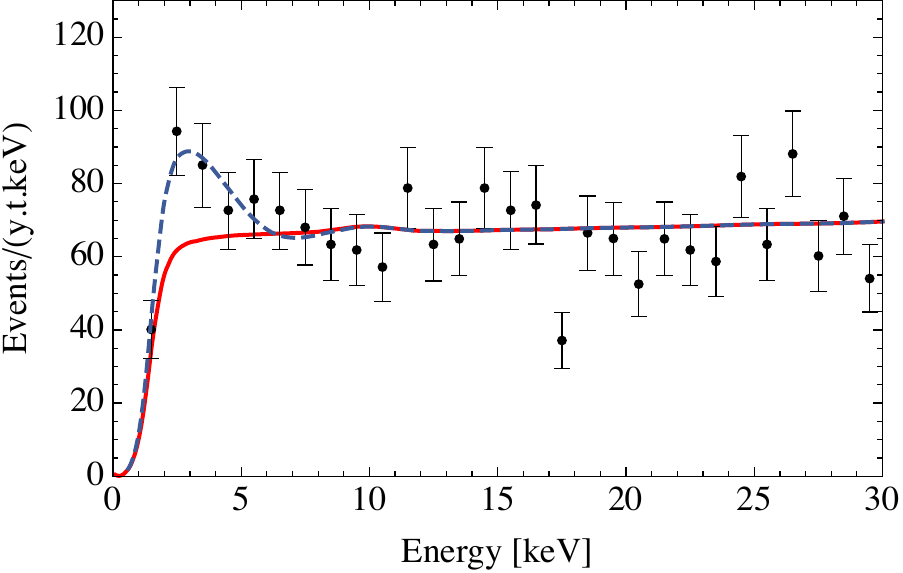} 
\caption{The signal in Xenon1T from Rayleigh DM scattering for the best fit point, $m_\varphi=1.9$ GeV, $\Lambda/\sqrt{C_\gamma}=50$ MeV (blue dashed line) compared to the background only hypothesis (red). Xenon1T data points are indicated with black error bars. }
	\label{fig:best:fit}
\end{figure}

\begin{figure}[t]
 \includegraphics[width=6.cm]{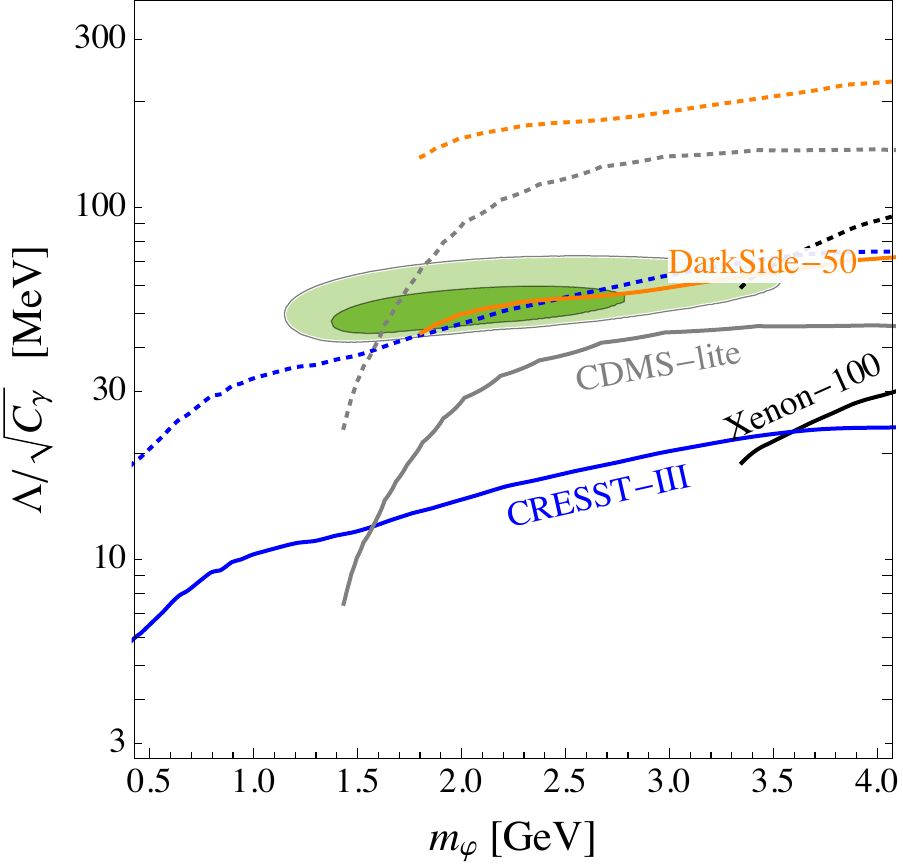} 
\caption{The preferred region in the plane  $m_\varphi$ vs. $\Lambda/\sqrt{C_\gamma}$ for the Rayleigh dark matter, assuming this is the total of DM relic abundance, $f_\varphi=1$, with $1\sigma(2\sigma)$ region that explains the Xenon1T anomaly shown with dark (light) green shading. The lower constraints on $\Lambda/\sqrt{C_\gamma}$ from one-loop induced  SI nuclear scattering, which suffer from large nonperturbative matrix element uncertainties, are denoted with blue (CRESST-III \cite{Abdelhameed:2019hmk}), gray (CDSM-lite  \cite{Agnese:2015nto}), orange (DarkSide-50 \cite{ Agnes:2018ves}), and black  (Xenon-100 \cite{Aprile:2016wwo}) lines for the nuclear coherence factors $\kappa=0.5, 0.05$ (dotted and solid, respectively).}
	\label{fig:best:region}
\end{figure}

We fit for the optimal Rayleigh DM signal, ignoring $E_{\rm NR}$ contributions, using a $\chi^2$ constructed from Xenon1T measurements in the recoil energy interval up to 30 keV, with the efficiency curve and the nominal background model given in Ref.~\cite{Aprile:2020tmw}. The best fit point has a significance of $3.3\sigma$ over the background only hypothesis, and is obtained for $m_\varphi=1.9$ GeV and $C_\gamma/\Lambda^2=1/\big(f_\varphi(50~{\rm MeV})^2\big)$, where $f_\varphi \equiv \Omega_{\varphi}/\Omega_{\rm DM}$ is the fraction of DM relic abundance that is due to $\varphi$. The comparison with Xenon1T data, depicted as black points with error bars, is given in Fig. \ref{fig:best:fit}. The signal due to Rayleigh DM (the Xenon1T background prediction) is shown with blue dashed (solid red) line. Since the Rayleigh DM signal is relatively wide, the energy smearing by the detector \cite{Aprile:2020yad,Aprile:2017aty} does not lead to any visible effect (to speed up the fit  smearing is not used in the $\chi^2$, nor in Fig. \ref{fig:best:fit}). Varying the DM mass and the effective scale $\Lambda/\sqrt{C_\gamma}$ gives the $1\sigma~(2\sigma)$ preferred regions, shown with dark (light) green shading in Fig. \ref{fig:best:region}. The DM mass in the range $\sim 1$ to $3.5$ GeV is preferred, on the border of the detection threshold for Xenon1T. The effective scale $\Lambda/\sqrt{C_\gamma}$ is in the range of ${\mathcal O}(50~{\rm MeV})$.

As we can see, the Rayleigh DM scattering describes the observed Xenon1T excess rather well. Next, we explore whether the low effective New Physics (NP) scale $\Lambda$ can be phenomenologically viable, starting with the induced SI nuclear recoil scattering. 

{\bf Spin-independent nuclear scattering.}
At one loop the Rayleigh operator generates SI scattering on the nuclei, $\varphi N\to \varphi N$, through the two photon exchange diagram, see Fig. \ref{fig:diagram} (right). This contribution is dominated by nuclear scales and is thus described by a nonperturbative matrix element. For a spin-$1/2$ nucleus $N$ the matrix element is defined as
\beq
\langle f| \big(\varphi \varphi\big)F_{\mu\nu} F^{\mu\nu} |i\rangle=\frac{\alpha Z^2}{4\pi} \widetilde Q_0 \langle f| \big(\varphi \varphi\big) \bar u_N u_N |i \rangle.
\eeq
For spin-0 nuclei we substitute $\bar u_N u_N\to 2m_N$ in the above expression. The initial and final $|\varphi\rangle |N\rangle$ states are shortened as $|i\rangle, |f\rangle$. The prefactor $\alpha Z^2/(4\pi)$ is based on naive dimensional analysis, assuming coherent scattering of two photons on the total charge of the nucleus. The nonperturbative parameter $\widetilde Q_0$ has a dimension of GeV and is expected to be parametrically of the inverse size of the corresponding nucleus, $\widetilde Q_0=\kappa /\sqrt{\langle r^2\rangle}$, where $ \sqrt{\langle r^2\rangle}$ is the charge radius of the nucleus. In the numerical analysis we use two values for the coherence factor, $\kappa=0.5 $ and $\kappa=0.05$, to show the uncertainties related to this otherwise completely unknown matrix element. The perturbative two-photon exchange model gives larger estimates for $\tilde Q_0$~\cite{Weiner:2012cb,Frandsen:2012db}. We also expect the Rayleigh operator to mix into dimension 5 DM--scalar-quark-current operators at one loop, leading to destructive interference in direct detection rate \cite{Frandsen:2012db}. This highlights the uncertainties surrounding the estimates of $\tilde Q_0$ nuclear matrix elements. Also, for potentially important contributions from two-body currents see Ref. \cite{Ovanesyan:2014fha}.

The SI $\varphi N\to \varphi N$ scattering cross section is $\sigma_N=A^2 (\mu_{\varphi N}^2/\mu_{\varphi n}^2) \sigma_n$, where 
\beq
\sigma_n=\frac{1}{4 \pi} \biggr(\frac{\alpha}{12\pi}\frac{C_\gamma}{\Lambda^2}\biggr)^2 \biggr(\frac{\alpha Z^2}{2\pi} \tilde Q_0\biggr)^2 \frac{\mu_{\varphi n}^2}{\mu_{\varphi N}^2}\frac{1}{A^2},
\eeq
is the SI  cross section for a single nucleon on which bounds are quoted by the direct detection experiments, see, e.g, \cite{Belanger:2008sj,Kopp:2009qt}. 
Here, $A$ is the atomic number, and $\mu_{\varphi n(N)}$ the reduced mass of the DM--nucleon (nucleus) system. While the $\varphi N\to \varphi N$ scattering 
cross section is loop suppressed, it is still much larger than the $\varphi N\to \varphi N \gamma$ cross section, which has the phase space suppression due to the 
extra particle in the final state, and, more importantly, extra suppression due to small available recoil energies. The $\varphi N\to \varphi N$ cross section is 
enhanced by a much larger dimensionful quantity $\widetilde Q_0^2$.

A number of direct detection experiments were able to probe the low mass DM region using nuclear recoils with low thresholds. The most important constraints for the case of Rayleigh DM are shown in Fig. \ref{fig:best:region}, with blue lines denoting CRESST-III \cite{Abdelhameed:2019hmk}, gray lines CDMS-lite \cite{Agnese:2015nto}, orange  lines DarkSide-50 \cite{ Agnes:2018ves}, and black lines the Xenon-100 low mass dark matter search \cite{Aprile:2016wwo}, where the dotted (solid) lines correspond to coherence factors $\kappa=0.5 (0.05)$. The regions below the lines are excluded for the assumed inputs. 
Even though one cannot draw definitive conclusions due to the large uncertainties, it is still quite likely that for $m_\varphi\gtrsim1.8$ GeV the region preferred by Xenon1T anomaly is excluded by the SI nucleon scattering search by Xenon-100,  as this would require a significantly suppressed nuclear nonperturbative matrix element. For lower masses, however, the region is most probably allowed since exclusions would require enhanced nonperturbative matrix elements instead. 

The one-loop two photon exchange also induces scattering of DM on electrons (for collection of present experimental results see \cite{Barak:2020fql}). However, these cross sections are parametrically smaller, suppressed by $m_e^2$ and do not lead to relevant constrains on $\Lambda/\sqrt{C_\gamma}$. 

{\bf Secluded DM.} The relatively low effective scale in the Rayleigh operator, $\Lambda/\sqrt{C_\gamma}\sim {\mathcal O}(50~{\rm MeV})$ can be easily realized if DM is secluded, i.e., if it does not directly couple to  the visible matter but rather through a mediator. 
We consider a simple model where the interaction with photon is mediated through a light (pseudo)scalar $a$ with mass $m_a\sim{\cal O}(1-10~{\rm MeV})$. The relevant interaction terms are
%
\beq
\begin{split}\label{eq:La}
{\cal L}_{a}\ \supset \ &\mu_\varphi \varphi^2 a+ \frac{\alpha}{12\pi }\frac{a}{\Lambda_{\rm UV}}\big(C_{a\gamma} FF
+\widetilde C_{a\gamma}F \tilde F\big),
\end{split}
\eeq
where $FF=F_{\mu\nu} F^{\mu\nu}$, $F\tilde F= F_{\mu\nu} \tilde F^{\mu\nu}$.
For momenta exchanges below $m_a$, which is the case for Xenon1T anomalous events,  the light scalar $a$ can be integrated out, resulting in the CP even Rayleigh operator in \eqref{eq:Rayleigh:scalar} with 
\beq\label{eq:translation}
\frac{C_\gamma}{\Lambda^2}=\frac{C_{a\gamma}}{\Lambda_{\rm UV}}\frac{\mu_\varphi}{m_a^2},
\eeq
and similarly for the CP odd coupling, with $C_\gamma \to \tilde C_\gamma$, $C_{a\gamma}\to \tilde C_{a\gamma}$ replacements. 

The tree level exchange of $a$ leads to a large self-interaction cross section for $\varphi \varphi\to \varphi\varphi$ scattering, well above the QCD cross section for $m_\varphi \sim {\mathcal O}({\rm GeV})$. This is excluded by astrophysical observations, if $\varphi$ is the dominant contribution to the DM relic density, but is allowed if $\varphi$ is a subdominant component, for $f_\varphi$ of a few tens of percent (comparable with the baryonic energy density) \cite{Fan:2013yva}.

The best fit point of the Xenon1T anomaly is obtained for 
\beq
\frac{\Lambda_{\rm UV}}{C_{a\gamma}}=1~{\rm TeV}\biggr(\frac{f_\varphi}{0.2}\biggr) \biggr(\frac{1{\rm~MeV}}{m_a}\biggr)^2 \biggr(\frac{\mu_\varphi}{2{\rm~GeV}}\biggr),
\eeq
and can thus be due to $a$ coupling to TeV scale particles, e.g., vector-like fermions, that carry electroweak charges and induce dimension 5 couplings in \eqref{eq:La}. Another attractive option is that $a$ is a pseudo Nambu-Goldstone boston of a global symmetry that is broken at the TeV scale and is anomalous with respect to $U(1)_{\rm em}$ in which case the $a\tilde F F$ term is induced. This has the benefit that the shift symmetry protects the mass of $a$. This symmetry is broken via couplings to DM which would induce $m_a\sim{\mathcal O}(1-10{\rm~MeV})$.

{\bf Indirect DM constraints}. In the secluded DM model there are  two types of tree level processes that give gamma ray line signals from DM annihilations to photons. The first is $s$ channel $\varphi\varphi$ annihilation from $a$ exchange, $\varphi\varphi\to a^*\to \gamma\gamma$, where $a$ is far off-shell since $m_a\ll m_\varphi$. This gives a gamma ray line at $m_\varphi$ with the annihilation cross section 
$
( \sigma v)_{\varphi\varphi\to 2\gamma}=\big[{\mu_\varphi \alpha C_{a\g}}/({12\pi\Lambda_{\rm UV}})\big]^2/({4 \pi m_\varphi^2}).
$

The $\varphi\varphi \to a a$ annihilation, where $a$ decays to two photons, also gives in the limit  $m_a\ll m_\varphi$ a line-shaped gamma-ray signal but at $m_\varphi/2$. The relative width of the gamma-ray line is given by $m_a/m_\varphi$ and is in our case small, $\sim 10^{-3}$. The $\varphi\varphi \to a a$ annihilation cross section, induced by the trilinear coupling, is
$
(\sigma v)_{\varphi\varphi\to 2a}={\mu_\varphi^4}/({8\pi m_\varphi^6}),
$
and is in general large, barring possible cancellations with the quartic contributions in ${\cal L}_a$. If $a$ decays predominantly to two photons this would lead to an untenably large signal in gamma ray flux in the sky. 
We thus assume that $a$ decays predominantly to either neutrinos or other invisible states, such that $Br(a\to \gamma\gamma)$ is below ${\mathcal O}(10^{-7})$, in which case the bounds from gamma ray lines are avoided. 

\begin{figure}[t]
 \includegraphics[width=6.cm]{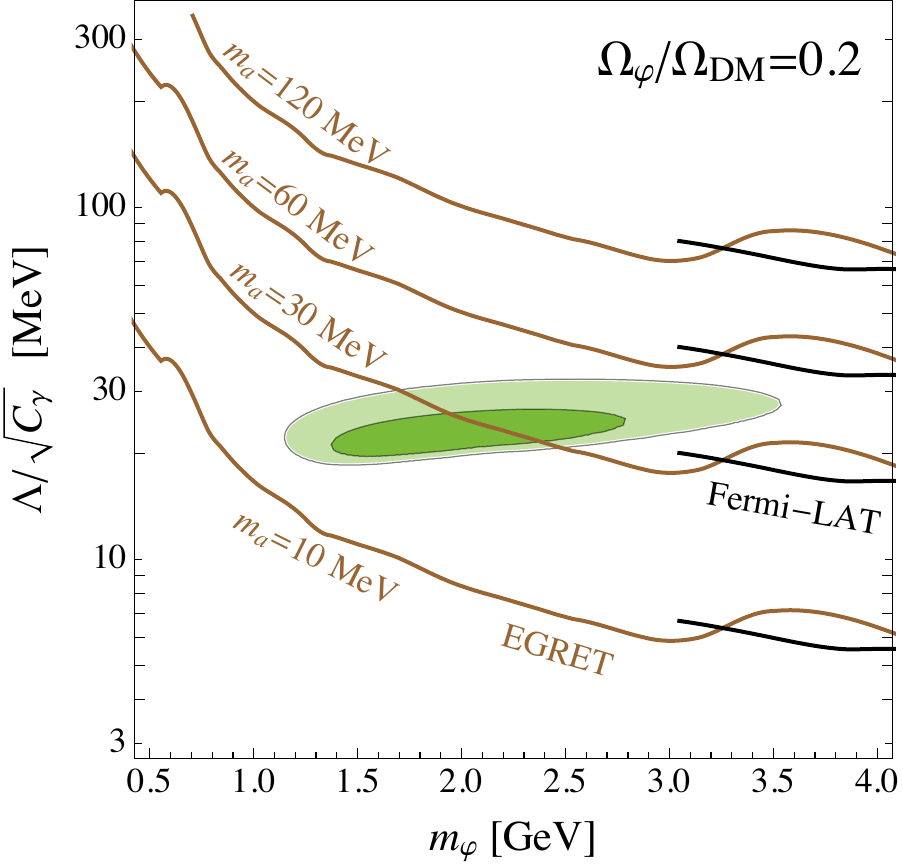} 
\caption{The region preferred by the Xenon1T anomaly is shown with green shading, as in Fig.~\ref{fig:best:region}, while the brown solid lines shown constrains from gamma ray line searches from dark matter annihilation in the galactic center, due to EGRET \cite{Pullen:2006sy} (brown) and Fermi-LAT\cite{Anderson_2016} (black) data, for several values of mediator, as indicated. }
	\label{fig:indirect}
\end{figure}

The constraints from $\varphi \varphi \to 2\gamma$ are shown in Fig.~\ref{fig:indirect} for  $f_\varphi=0.2$ (for smaller values of $f_\varphi$ the indirect bounds become less important). For easier comparison with Fig.~\ref{fig:best:region} we translate, using Eq. \eqref{eq:translation}, the constraints  to the upper bounds on the effective NP scale of the Rayleigh operator,  $\Lambda/C_\g$, choosing several representative values of $m_a=\{2,10,30,80\}$ MeV. The brown (black) lines show the corresponding 90\% CL limits from gamma-ray emissions in the Galactic Center as observed by EGRET  \cite{Pullen:2006sy} (Fermi-LAT \cite{Anderson_2016}). The Fermi-LAT constraints start at about $m_\varphi=3$ GeV and overlap with EGRET, for given value of $m_a$. 
The region preferred by the Xenon1T anomaly is not constrained, if the mediator is lighter  than about $m_a\sim10~{\rm MeV}$.

Since the mediator decays invisibly the direct constraints on the production of $a$ from colliders or in beam-dumps are significantly weakened. Interestingly, even for a pseudoscalar of a mass of a few 10's of MeV that couples exclusively to photons, the bounds are quite weak, see, e.g., Ref. \cite{Beacham:2019nyx}. 

{\bf Discussion and conclusions.} The intriguing ``electron recoil'' excess events in Xenon1T can be due to non-relativistic DM scattering off nuclei, if the scattering is induced by Rayleigh operators, since the Xenon1T detector does not discriminate between electrons and photons.
In our analysis we neglected the small nuclear recoil contribution, which we believe to be an excellent approximation. It would be useful that this is checked by a detailed detector response simulation. 

The Rayleigh DM model that explains the Xenon1T anomaly faces, unsurprisingly, severe constraints from indirect detection and one-loop induced SI scattering. We have showed that viable parameter space exists for the case of MeV scale mediators. There are several uncertainties that enter the discussion. First of all, due to uncertain nonperturbative nuclear matrix elements 
one cannot draw definite conclusions to what extent the Rayleigh DM model that explains Xenon1T is constrained by bounds on spin-independent scattering on nuclei. 
Furthermore, the signal arises from parts of the DM velocity distribution that are relatively close to the escape velocity, and is thus subject to enhanced uncertainties in the DM halo velocity distributions. It would be interesting to revisit these issues as well as other effects related to our setup, such as the modifications of the Xenon1T signal, if the mediator is lighter than the EFT limit. 

The suggested Rayleigh DM model can be probed experimentally, by improving the experimental bounds on low mass DM searches from nuclear recoils, by searching for neutrino interactions with the MeV scale mediator, or for the production of the weakly-coupled mediator in collider experiments. 

{\bf Acknowledgements:} We thank Ranny Budnik, Felix Kahlhoefer, and Diego Redigolo for useful discussions. JZ and MT acknowledge support in part by the DOE grant de-sc0011784.
GP and AAP were supported in part by the DOE grant de-sc0007983. GP was also supported in part by a Career Development 
Chair award from Wayne State University.

\bibliographystyle{h-physrev}
\bibliography{DMbiblio}

\end{document}